\documentclass[reprint,superscriptaddress,amsmath,amssymb,aps,prx]{revtex4-2}
\usepackage{graphicx}
\usepackage{dcolumn}
\usepackage{bm}
\usepackage{qcircuit}
\usepackage{braket}
\usepackage{color}
\usepackage{amsthm}
\usepackage{chemarrow}
\usepackage{overpic}
\usepackage{diagbox}
\usepackage{braket}
\usepackage{subfigure}
\usepackage{makecell}
\newtheorem{theorem}{Theorem}
\newtheorem{definition}{Definition}

\usepackage{chngcntr}
\usepackage[colorlinks=true,linkcolor=red,bookmarks=true,breaklinks=true]{hyperref}
\usepackage{caption}
\usepackage{algorithm}
\usepackage{algorithmic}

\begin{document}
\newcommand{\Tr}{\mathrm{Tr}}
\newcommand{\re}{\text{Re}}
\newcommand{\im}{\text{Im}}
\newcommand{\szx}[1]{\textcolor{blue}{(szx: #1)}}
\title{Pauli quantum computing: $I$ as $|0\rangle$ and $X$ as $|1\rangle$}
\author{Zhong-Xia Shang}
\email{shangzx@hku.hk}
\affiliation{HK Institute of Quantum Science $\&$ Technology, The University of Hong Kong, Hong Kong, China}
\affiliation{QICI Quantum Information and Computation Initiative, Department of Computer Science,
The University of Hong Kong, Hong Kong, China}
\begin{abstract}
We propose a new quantum computing formalism named Pauli quantum computing. In this formalism, we use the Pauli basis $I$ and $X$ on the non-diagonal blocks of density matrices to encode information and treat them as the computational basis $|0\rangle$ and $|1\rangle$ in standard quantum computing. There are significant differences between Pauli quantum computing and standard quantum computing from the achievable operations to the meaning of measurements, resulting in novel features and comparative advantages for certain tasks. We will give three examples in particular. First, we show how to design Lindbladians to realize imaginary time evolutions and prepare stabilizer ground states in Pauli quantum computing. These stabilizer states can characterize the coherence in the steady subspace of Lindbladians. Second, for quantum amplitudes of the form $\langle +|^{\otimes n}U|0\rangle^{\otimes n}$ with $U$ composed of $\{H,S,T,\text{CNOT}\}$, as long as the number of Hadamard gates in the unitary circuit $U$ is sub-linear $\mathit{o}(n)$, the gate (time) complexity of estimating such amplitudes using Pauli quantum computing formalism can be exponentially reduced compared with the standard formalism ($\mathcal{O}(\epsilon^{-1})$ to $\mathcal{O}(2^{-(n-\mathit{o}(n))/2}\epsilon^{-1})$). Third, given access to a searching oracle under the Pauli encoding picture manifested as a quantum channel, which mimics the phase oracle in Grover's algorithm, the searching problem can be solved with $\mathcal{O}(n)$ scaling for the query complexity and $\mathcal{O}(\text{poly}(n))$ scaling for the time complexity. While so, how to construct such an oracle is highly non-trivial and unlikely efficient due to the hardness of the problem. 
  
\end{abstract}
\maketitle

\section{Introduction}
Quantum computers can exploit quantum mechanical phenomena such as superposition and interference to give speedup over classical computers for solving certain problems \cite{feynman2018simulating,shor1999polynomial,grover1996fast,nielsen2010quantum,dalzell2023quantum}. What we are doing with a quantum computer is basically that we start from an initial state, implement a unitary quantum circuit, and then measure the output state (Born’s rule). The basic unit of information in standard quantum computing is a qubit, a 2-dimensional quantum system and we use its two basis states $|0\rangle$ and $|1\rangle$ to play the roles of $0$ and $1$ in a classical bit. 

While this $0-|0\rangle$ and $1-|1\rangle$ correspondence seems to come naturally, in this work, we ask whether there are other choices for encoding information leading to new quantum computing formalism. What we mean by this is not simply running a unitary transformation to $|0\rangle$ and $|1\rangle$ and using the transformed states to encode $0$ and $1$ such as quantum error correction codes but other fundamentally different choices. The particular choice we are interested in in this work is the basis of density matrices. Given that we have used the basis $|0\rangle$ and $|1\rangle$ in the space of wave function (pure state) to encode information and do the quantum computing, we should also try the basis in the space of density matrices. Quantum applications based on $|i\rangle\langle j|$ basis of density matrices have been explored in recent works \cite{shang2024polynomial,shang2024unconditionally,shang2024estimating}. Here, we instead consider the Pauli basis. Pauli operators have played numerous important roles in the theory of quantum computing from the stabilizer quantum error correction codes \cite{gottesman1997stabilizer} to the magic of quantum states \cite{leone2022stabilizer} and the classical simulation of (noisy) quantum circuits \cite{gottesman1998heisenberg,aaronson2004improved,aharonov2023polynomial,schuster2024polynomial}. In this work, we want to give them another job: we propose the Pauli quantum computing (PQC) where we use the Pauli operators $I$ and $X$ in density matrices to encode information and take the roles of $|0\rangle$ and $|1\rangle$ in standard quantum computing (SQC). 

While the benefit of PQC is not obvious, we have the observation that under this new formalism, the rules of state preparation, operation, and measurement in SQC are vastly changed, and thus, we can expect some novel features and comparative advantages to emerge. The rest of the paper is composed of two parts. In the first part, we introduce the framework of PQC formalism. We will justify the meaning of treating $I$ as $|0\rangle$ and $X$ as $|1\rangle$ and show how to get rid of the Hermitian and positive semi-definite restrictions of density matrices by harnessing their non-diagonal blocks. We will also give a systematic way on how to construct desirable operations by quantum channels and introduce how to do measurements for amplitudes and operator expectation values. 

In the second part, we will give three examples of PQC to help understand its various aspects and differences from SQC. In the first example, we show how to prepare stabilizer ground states \cite{gottesman1997stabilizer} under PQC by the imaginary time evolution realized by a class of Lindbladians \cite{lindblad1976generators,gorini1976completely}, a model for open quantum systems. We name these ground states the density matrix stabilizer coherence. The second example is the complexity reduction for estimating certain quantum amplitudes under PQC. Specifically, PQC can give exponential improvements in the gate (time) complexity over SQC when estimating amplitudes $\langle +|^{\otimes n}U|0\rangle^{\otimes n}$ with $U$ containing $\mathit{o}(n)$ Hadamard gates. The last example is quantum searching under PQC. We will show that if we are able to have access to a quantum channel that is a Pauli searching oracle mimicking the phase oracle in Grover's algorithm \cite{grover1996fast} in SQC, the search problem can be solved efficiently with only $\mathcal{O}(n)$ queries and $\mathcal{O}(\text{poly}(n))$ time. However, we want to emphasize that this has nothing to do with $\text{NP}\in \text{BQP}$ since how to efficiently construct such an oracle is unknown.

\section{Pauli quantum computing}

\subsection{Vectorization and matrixization}
To facilitate later discussions, we first introduce the notions of vectorization mapping and matrixization mapping.
\begin{definition}[Vectorization and Matrixization]
Given a $n$-qubit matrix $O=\sum_{ij}o_{ij}|i\rangle\langle j|$, the vectorization mapping $\mathcal{V}$ has the operation: $$\mathcal{V}[O]=\ket{O\rangle}=(O\otimes I_n) |\Omega\rangle=\sum_{ij}o_{ij}|i\rangle|j\rangle,$$
where $|\Omega\rangle=2^{-n/2}\sum_i |i\rangle|i\rangle$ is the $2n$-qubit maximally entangled Bell state. The reverse mapping of the vectorization is called the matrixization $\mathcal{M}$ which has the operation:
$$\mathcal{M}[\ket{O\rangle}]=O.$$
\end{definition}
\noindent In this work, we use the symbol $\ket{\cdot\rangle}$ when the vector is not normalized. Based on the vectorization mapping, we have the Pauli-Bell correspondence:
\begin{eqnarray}\label{PBC}
I=\begin{pmatrix}1& 0\\0&1\end{pmatrix}&&\xrightarrow[]{\mathcal{V}}\sqrt{2}|\Phi^+\rangle=|0\rangle|0\rangle+|1\rangle|1\rangle\nonumber,\\
Z=\begin{pmatrix}1& 0\\0&-1\end{pmatrix}&&\xrightarrow[]{\mathcal{V}}\sqrt{2}|\Phi^-\rangle=|0\rangle|0\rangle-|1\rangle|1\rangle\nonumber,\\X=\begin{pmatrix}0& 1\\1&0\end{pmatrix}
&&\xrightarrow[]{\mathcal{V}}\sqrt{2}|\Psi^+\rangle=|0\rangle|1\rangle+|1\rangle|0\rangle\nonumber,\\
Y=\begin{pmatrix}0& -i\\i&0\end{pmatrix}&&\xrightarrow[]{\mathcal{V}}-i\sqrt{2}|\Psi^-\rangle=|0\rangle|1\rangle-|1\rangle|0\rangle.
\end{eqnarray}

\begin{figure*}[htbp]
\centering
\includegraphics[width=0.85\textwidth]{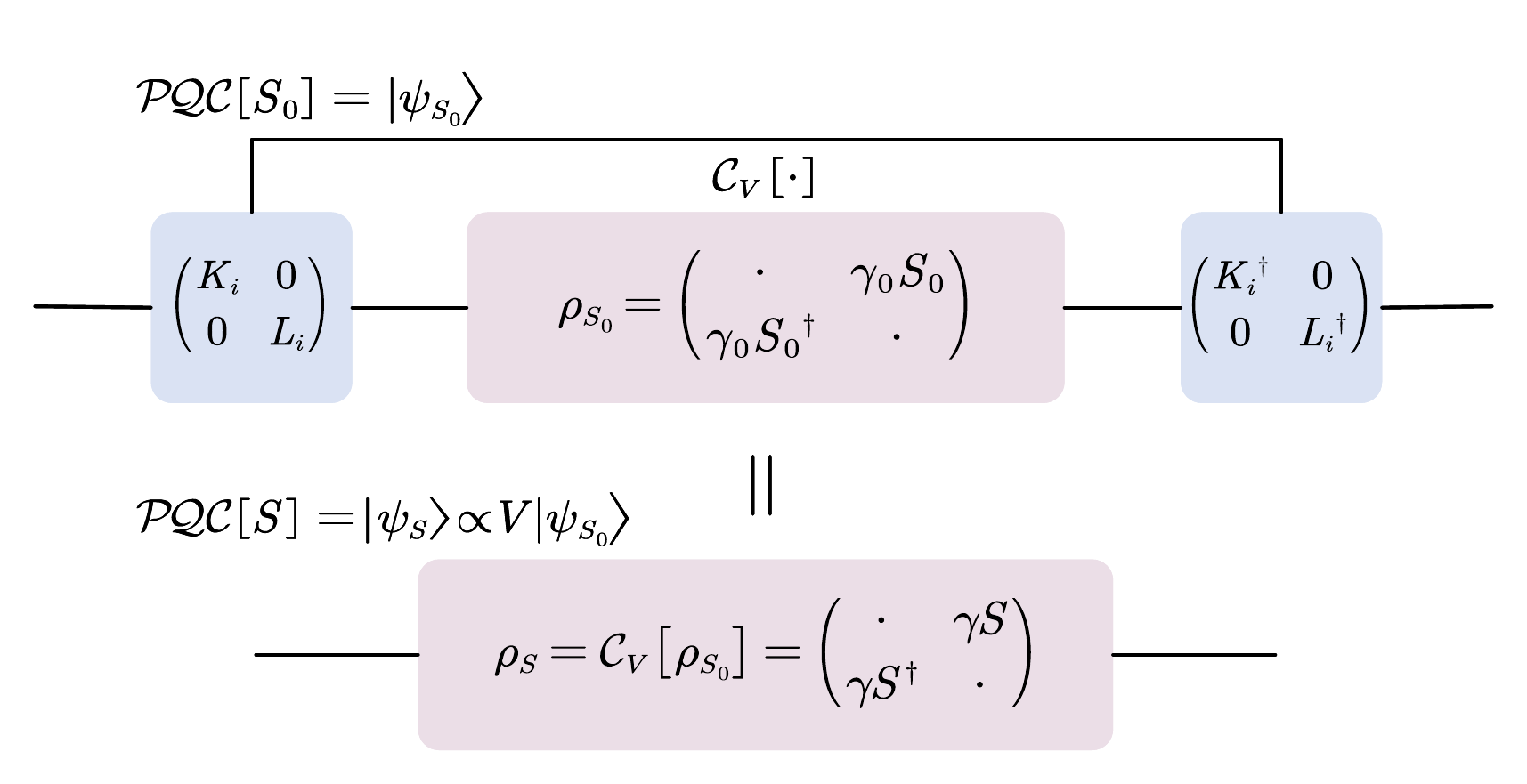}
\caption{Framework of Pauli quantum computing. In Pauli quantum computing, we treat $I$ as $|0\rangle$ and $X$ as $|1\rangle$. By this treatment, we are able to encode the information of a pure state into a matrix which is further embedded into the non-diagonal block of a density matrix. Starting from an initial density matrix $\rho_{S_{0}}$ that encodes the information of a pure state $|\psi_{S_{0}}\rangle$, we can implement a quantum channel with a special form to generate an output density matrix $\rho_{S}=\mathcal{C}_V[\rho_{S}]$ which encodes the information of $V|\psi_{S_{0}}\rangle$. $V$ is not necessarily unitary. $\rho_{S}$ can be further measured to extract information.\label{fig1}}
\end{figure*}

\subsection{Non-diagonal density matrix encoding and channel block encoding}
Given a $n$-qubit density matrix $\rho$, we have the Pauli decomposition. The philosophy of this work is that we want to treat the Pauli operators as the computational basis and the decomposition coefficients as the amplitudes in SQC. However, this naive encoding has many restrictions on amplitudes due to the Hermiticity and positive semi-definiteness of density matrices. To get rid of such restrictions, we consider using non-diagonal blocks of density matrices to encode information, which we named Non-diagonal density matrix encoding (NDME):
\begin{definition}[Non-diagonal density matrix encoding (NDME) \cite{shang2024design,shang2024estimating}]
Given an $(1+n)$-qubit density matrix $\rho_O$, a $n$-qubit matrix $O$, and a positive number $\gamma$, $\rho_O$ is called $\gamma$-NDME of $O$ if $\rho_O$ satisfies:
$$(\langle 0|\otimes I_n) \rho_O (|1\rangle\otimes I_n)=\gamma O.$$
\end{definition}
We will call the first qubit the assistant qubit and the rest of the $n$-qubit system the encoding system. NDME gives the ability to encode an arbitrary matrix into the Hermitian and positive semi-definite density matrix. To be more clear, $\rho_O$ has the form:
\begin{eqnarray}
\rho_O=\begin{pmatrix}
\cdot & \gamma O \\
\gamma O^\dag & \cdot
\end{pmatrix}.
\end{eqnarray} 

Having the NDME, we can further consider operations we can implement to $O$. We define a special class of quantum channels $\mathcal{C}[\cdot]$ with the form:
\begin{equation}\label{channel}
\mathcal{C}[\rho]=\sum_i \begin{pmatrix}
K_i & 0\\
0& L_i
\end{pmatrix}\rho \begin{pmatrix}
K_i^\dag & 0\\
0& L_i^\dag
\end{pmatrix},
\end{equation}
where $\{K_i\}$ and $\{L_i\}$ are two sets of Kraus operators with $\sum_i K_i^\dag K_i=\sum_i L_i^\dag L_i=I_n$, following the fact that quantum channels are completely positive trace-preserving (CPTP) maps. Acting this channel to $\rho_O$ gives:
\begin{equation}
\mathcal{C}[\rho_{O}]=\begin{pmatrix}
\cdot & \gamma\sum_i K_iOL_i^\dag \\
\gamma\sum_i L_i O^\dag K_i^\dag & \cdot
\end{pmatrix}.
\end{equation}
Focus on the upper-right block and do the vectorization mapping, we have:
\begin{equation}
\mathcal{V}[\gamma\sum_i K_iOL_i^\dag]=\gamma\left(\sum_i K_i\otimes L_i^*\right)\ket{O\rangle}.
\end{equation}
Thus, by adjusting $\{K_i\}$ and $\{L_i\}$, we can realize desired actions to $\ket{O\rangle}$ in the vectorization picture, which invokes us to propose the channel block encoding (CBE):
\begin{definition}[Channel block encoding (CBE) \cite{shang2024design,shang2024estimating}]
Given a $2n$-qubit operator $Q$, a $1+n$-qubit quantum channel $\mathcal{C}[\cdot]$ of the form Eq. \ref{channel}, and a positive number $\eta$, $\mathcal{C}[\cdot]$ is called a $\eta$-CBE of $Q$ if:
$$\sum_i K_i\otimes L_i^*=\eta Q.$$
\end{definition}

\subsection{$I$ as $|0\rangle$ and $X$ as $|1\rangle$}
Pauli quantum computing can be better understood from the vectorization picture. First, we consider a simple Bell circuit $U_B$:
\begin{equation}
U_B=\Qcircuit @C=1em @R=.7em {
&\qw &\ctrl{1}  & \gate{H} &\qw \\
&\qw &\targ  &  \qw &\qw}=(H\otimes I)\text{CNOT},
\end{equation}
which has the transformation: $U_B|\Phi^+\rangle=|0\rangle|0\rangle$ and $U_B|\Psi^+\rangle=|0\rangle|1\rangle$. Second, according to Eq. \ref{PBC}, we know that the vectorization representation of $I$ and $X$ are $|\Phi^+\rangle$ and $|\Psi^+\rangle$ respectively. These two facts enable us to construct a mapping $\mathcal{PQC}[\cdot]$ from $I$ to $|0\rangle$ and $X$ to $|1\rangle$ by focusing on the second qubit:
\begin{eqnarray}\label{ix01}
\mathcal{PQC}[I]=(\langle 0|\otimes I )U_B\mathcal{V}[I]&&=\sqrt{2}|0\rangle,\nonumber\\
\mathcal{PQC}[X]=(\langle 0|\otimes I )U_B\mathcal{V}[X]&&=\sqrt{2}|1\rangle.
\end{eqnarray}
This mapping can be easily generalized to $n$-qubit Pauli to $2n$-qubit Bell cases. In the following, under the vectorization picture, we will arrange the $n$-qubit system where Hadamard gates in $U_B$ live in the front and the rest $n$-qubit system behind. We want to emphasize here that the mapping $\mathcal{PQC}[\cdot]$ doesn't have an operational meaning and the introduction of this mapping only aims to showcase and justify how we can \textbf{treat} $I$ as $|0\rangle$ and $X$ as $|1\rangle$.

Now, starting from a $n$-qubit matrix $S$ whose Pauli decomposition contains only Pauli operators composed of $I$ and $X$:
\begin{equation}\label{s}
S=2^{-n/2}\sum_\alpha c_\alpha Q_\alpha,
\end{equation}
satisfying $\sum_i |c_i|^2=1$, where the index $\alpha=\alpha_1...\alpha_n $ are $n$-bit binary strings to represent Pauli operators: when $\alpha_j=0$, the Pauli operator on the $j$th qubit is $I$, when $\alpha_j=1$, the Pauli operator on the $j$th qubit is $X$, the mapping Eq. \ref{ix01} transforms $S$ to:
\begin{equation}
\mathcal{PQC}[S]=\sum_\alpha c_\alpha |\alpha\rangle=|\psi_S\rangle.
\end{equation}
When a $1+n$-qubit density matrix $\rho_S$ is $\gamma$-NDME of $S$, $\rho_S$ under the PQC framework encodes the same information as the pure state $|\psi_S\rangle$ in SQC. For example, $\rho_S=|+\rangle\langle +|^{\otimes (n+1)}$ encodes $|\psi_S\rangle=|+\rangle^{\otimes n}$.

\subsection{Operations}

We can consider the action of the channel Eq. \ref{channel} to $\rho_S$, focusing on the upper-right block, we have:
\begin{eqnarray}
&&U_B^{\otimes n}\mathcal{V}[\gamma\sum_i K_iSL_i^\dag] =\nonumber\\&&\gamma U_B^{\otimes n} \left(\sum_i K_i\otimes L_i^* \right)U_B^{\dag\otimes n} |0\rangle^{\otimes n} |\psi_S\rangle.
\end{eqnarray}
If we want to implement an operator $V$ to $|\psi_S\rangle$, we require a CBE:
\begin{eqnarray}\label{po}
\sum_i K_i\otimes L_i^*=\eta U_B^{\dag\otimes n} (F_0\otimes V)U_B^{\otimes n},
\end{eqnarray}
where $\eta$ is a normalization factor and $F_0$ satisfies $F_0|0\rangle^{\otimes n}=|0\rangle^{\otimes n}$. It is important to note that $V$ is not necessarily unitary, in fact, it has been shown in Ref. \cite{shang2024estimating} that if we don't have requirements on $\eta$, for arbitrary $V$, we can always find a corresponding quantum channel $\mathcal{C}_V[\cdot]$ of the form Eq. \ref{channel} that satisfy Eq. \ref{po}. We give concrete optimal (the largest $\eta$) constructions of $\mathcal{C}_V[\cdot]$ with $V$ elementary quantum gates in Appendix \ref{A1}. We also explore realizing imaginary time evolutions by $\mathcal{C}_V[\cdot]$ in Section \ref{example1}.

\subsection{Measurement}
Given the density matrix $\rho_S$, here we show how to extract its information. We consider two tasks: estimating operator expectation values and estimating amplitudes.

For operator expectation values of the form $\langle \psi_S| P|\psi_S\rangle$ with $P$ a $2n$-qubit Pauli operator, we can first prepare the density matrix $\rho_{S_1}=\mathcal{C}_P[\rho_S]$ which is $\gamma\eta$-NDME of $S_1$ with $\mathcal{PQC}[S_1]=P|\psi_S\rangle$. For Pauli operators, we can realize $\mathcal{C}_P[\cdot]$ with $\eta=1$ (see Appendix \ref{A1}). Since we have $\langle \psi_S|P|\psi_S\rangle=\Tr(S^\dag S_1)$, we can estimate the expectation value by the following relation:
\begin{eqnarray}\label{expect}
\Tr(|01\rangle\langle 10|
\otimes \text{SWAP}_n (\rho_S\otimes \rho_{S_1}))=\gamma^{2}\langle\psi_S|P|\psi_S\rangle,\nonumber\\
\end{eqnarray}
where $|01\rangle\langle 10|$ is in two assistant qubits and $\text{SWAP}_n$ swaps the two encoding systems. From the operational perspective, we can construct a unitary operator as a block encoding \cite{low2019hamiltonian} of $|01\rangle\langle 10|
\otimes \text{SWAP}_n$, then the left-hand side of Eq. \ref{expect} can be estimated by Hadamard test \cite{cleve1998quantum}. If we further have the purification of $\rho_S$ and $ \rho_{S_1}$, the amplitude estimation algorithms \cite{brassard2002quantum,aaronson2020quantum,grinko2021iterative} can be adopted to acquire the Heisenberg limit.

For amplitudes of the form $\langle \alpha |\psi_S\rangle$ with $|\alpha\rangle$ a computational basis, we have the relation:
\begin{eqnarray}\label{ampp}
\Tr(X\otimes Q_\alpha \rho_S)&&=2^{n/2+1}\gamma\re[\langle \alpha |\psi_S\rangle]= 2^{n/2+1}\gamma\re[c_\alpha],\nonumber\\
\Tr(Y\otimes Q_\alpha \rho_S)&&=2^{n/2+1}\gamma\im[\langle \alpha |\psi_S\rangle]= 2^{n/2+1}\gamma\im[c_\alpha].\nonumber\\
\end{eqnarray}
Therefore, the $Q_\alpha$ Pauli expectation value of $\rho_S$ corresponds to the amplitudes of $|\psi_S\rangle$ on the computational basis $|\alpha\rangle$. Again, the Hadamard test or the amplitude estimation algorithms can be used to estimation the left-hand side of Eq. \ref{ampp}.

It is interesting to compare between Eq. \ref{expect} and Eq. \ref{ampp}. We will show later that the largest possible value of $\gamma$ is $1/2\leq 1$, therefore, in Eq. \ref{expect}, PQC shrinks the operator expectation values resulting in a larger complexity in estimation compared with SQC, in contrast, in Eq. \ref{ampp}, PQC enlarges the values of amplitudes when $2^{n/2+1}\gamma> 1$ resulting in a smaller complexity in estimation compared with SQC. We will elaborate on this point in Section \ref{example2}. 

\subsection{Brief summary}
We summarize the basic framework of PQC in Fig. \ref{fig1}. In standard quantum computing, we start from an initial state $|\psi_{S_{0}}\rangle$, then implement a unitary quantum circuit $U$ to generate an output state $U|\psi_{S_{0}}\rangle$ which is further measured to extract information. In comparison, in Pauli quantum computing, we use $I$ and $X$ to take the roles of $|0\rangle$ and $|1\rangle$ in SQC. We start from an initial density matrix $\rho_{S_{0}}$ which is $\gamma_{0}$-NDME of $S_{0}$ with $\mathcal{PQC}[S_{0}]=|\psi_{S_{0}}\rangle$, then we implement an operation $\eta V$ (\textbf{not necessarily unitary}) to $|\psi_{S_{0}}\rangle$ which is realized by a quantum channel $\mathcal{C}_V[\cdot]$ of the form Eq. \ref{channel} satisfying Eq. \ref{po}. The output density matrix $\rho_{S}$ therefore is $\gamma$-NDME of $S$ with $\mathcal{PQC}[S]=|\psi_{S}\rangle=V|\psi_{S_{0}}\rangle/\|V|\psi_{S_{0}}\rangle\|$ and $\gamma=\eta\gamma_{0}\|V|\psi_{S_{0}}\rangle\|$. $\rho_{S}$ is further measured to extract information such as amplitudes on the computational basis and the operator expectation values.

\section{Examples}

\subsection{Density matrix stabilizer coherence \label{example1}}
We consider preparing stabilizer ground states that are related to many important areas \cite{gottesman1997stabilizer,kitaev2003fault, aaronson2004improved,anshu2023nlts} in quantum information science by imaginary time evolution in PQC. This preparation is assisted by the Lindblad master equation (Lindbladian) \cite{lindblad1976generators,gorini1976completely}. Lindbladian is a general model for the Markovian dynamics of open quantum systems with the form:
\begin{eqnarray}\label{lind}
&&\frac{d\rho_{S,t}}{dt}=\mathcal{L}[\rho_{S,t}]\nonumber\\&&=-i[H,\rho_{S,t}]+
\sum_i \lambda_i\left(F_i\rho_{S,t} F_i^\dag-
\frac{1}{2}\{\rho_{S,t},F_i^\dag F_i\}\right),\nonumber\\
\end{eqnarray}
with $\lambda_i\geq 0$. Given a Hamiltonian $H_p=\sum_{i}\lambda_i P_i$ where $P_i$ are Pauli operators $\{\pm1\}\times\{I,X,Y,Z\}^{\otimes n}$, we can set a Lindbladian with the internal Hamiltonian $H=0$ and we require jump operators $F_i$ having the form: $F_i=\begin{pmatrix}
P_{1,i} & 0 \\
0 & P_{2,i}
\end{pmatrix}$ with $P_{1,i}$ and $P_{2,i}$ are Pauli operators $\{\pm1\}\times\{I,X,Y,Z\}^{\otimes n}$ such that Eq. \ref{po} is satisfied: $U_B^{\otimes n}(P_{1,i}\otimes P_{2,i}^*)U_B^{\dag\otimes n}=-I_n\otimes P_i$. Concrete constructions can be found in Appendix \ref{A1}. Since $F_i^\dag F_i=I\otimes I_n$, this Lindbladian therefore, under PQC, equivalently gives the imaginary time evolution: 
\begin{equation}\label{ite}
\frac{d\ket{\psi_{S,t}\rangle}}{dt}=\left(-H_p-\sum_{i}\lambda_i\right)\ket{\psi_{S,t}\rangle}.
\end{equation}

Eq. \ref{ite} can be used to prepare Gibbs states under PQC. For Hamiltonians with frustrations \cite{vojta2018frustration}, the ground energy is larger than $-\sum_{i}\lambda_i$, which leads to total decoherence on the non-diagonal block of $\rho_{S,t}$ i.e. $S=0$ at low temperature. In contrast, for frustration-free Hamiltonians, the Lindbladian evolution will only eliminate excited states and prepare (keep) the unnormalized ground state $\Pi_g|\psi_{S,0}\rangle$ with $\Pi_g$ the ground subspace projector. The norm of the prepared ground state only depends on the initial state. Due to Eq. \ref{ampp}, even if the norm is exponentially small, for amplitude estimation tasks, complexity might still be comparable with SQC. The surviving of ground state corresponds to the case where all $-P_i$ are elements of a stabilizer group $\mathcal{S}$ \cite{gottesman1997stabilizer}. Suppose the group $\mathcal{S}$ has $m$ independent generators, the resulting ground subspace dimension would be $2^{n-m}$. In the density matrix picture, this ground subspace can be used to characterize the steady-state properties of the Lindbladian: whenever $|\psi_O\rangle$ is in the ground subspace, $\Tr(\rho_{S,t} X\otimes O)$ is steady i.e. $d \Tr(\rho_{S,t} X\otimes O)/dt=0$. We call $X\otimes O$ the density matrix stabilizer coherence of the Lindbladian which may be of special interest in understanding unital noise. (We say the coherence of a density matrix is the deviation from the maximally mixed state.) Following the idea of this section, a similar argument should be easily generalized for $Y$, $Z$ coherence ($O$ is composed of $I$ and $X$ operators), and coherence in diagonal blocks of $\rho_{S,t}$.

\subsection{Estimation of amplitudes $\langle +|^{\otimes n}U|0\rangle^{\otimes n}$ \label{example2}}
Quantum amplitudes have played important roles in quantum computing as they can encode GapP problems that are foundations of quantum sampling advantage proposals \cite{hangleiter2023computational}. Here, we consider the advantages of PQC compared with SQC on estimating amplitudes $\langle \alpha|\psi_S\rangle$ with $V$ a unitary quantum circuit built by the universal gate set $\{H,S_g,T,\text{CNOT}\}$ (We use $S_g$ to denote the phase gate to avoid confusion with $S$ in Eq. \ref{s}.). Pauli quantum computing has a remarkable feature in Pauli measurements as shown in Eq. \ref{ampp}: the Pauli expectation values of $\rho_S$ correspond to the amplitudes of $|\psi_S\rangle$ on the computational basis. It is important to notice the factor $2^{n/2+1}\gamma$. Since both the Pauli operators and the projectors $|\alpha\rangle\langle \alpha|$ have spectra within the range $[-1,1]$, estimating the left-hand side of Eq. \ref{po} under Pauli quantum computing framework and estimating $\langle \alpha|\psi_S\rangle$ in standard quantum computing has comparable complexity for the same additive error. However, due to the existence of the factor $2^{n/2+1}\gamma$, when this factor is large, we are able to use Pauli quantum computing to significantly reduce the complexity of estimating the amplitudes $\langle \alpha|\psi_S\rangle$.

Depending on the estimation techniques, we can either achieve a standard quantum limit estimation ($\epsilon^{-2}$) or achieve a Heisenberg limit ($\epsilon^{-1}$). Here, we focus on the Heisenberg limit estimation. Consider the purification $|P_S\rangle$ of $\rho_S$, we have:
\begin{eqnarray}\label{hle}
|\langle P_S|I_e\otimes (X\otimes Q_\alpha+I\otimes I_n) |P_S\rangle|&&=1+2^{n/2+1}\gamma \re[c_\alpha],\nonumber\\
|\langle P_S|I_e\otimes (Y\otimes Q_\alpha+I\otimes I_n) |P_S\rangle|&&=1+2^{n/2+1}\gamma \im[c_\alpha],\nonumber\\
\end{eqnarray}
where $I_e$ is the identity operator in the environment system. Since $(X\otimes Q_\alpha+I\otimes I_n)/2$ can be block-encoded into a unitary operator \cite{rall2020quantum}, the left-hand side of Eq. \ref{hle} can be viewed as amplitudes and thus can be estimated by amplitude estimation algorithms \cite{brassard2002quantum,aaronson2020quantum,grinko2021iterative}. The result is that if we want to estimate $\re[c_\alpha]=\re[\langle \alpha|\psi_S\rangle]$ to an additive error $\epsilon$, the query complexity on the preparation circuit (the unitary purification of $\mathcal{C}_V[\cdot]$) of $|P_S\rangle$ is $\mathcal{O}(2^{-n/2}\gamma^{-1}\epsilon^{-1})$. The same conclusion also applies to the imaginary part.

To make Pauli quantum computing have truly time-complexity advantages in estimating $\langle \alpha|\psi_S\rangle$, we require $2^{n/2+1}\gamma\gg 1$ and the gate complexity of $\mathcal{C}_V[\cdot]$ (as well as its unitary purification) is comparable with the gate complexity of $V$. The second requirement is obviously satisfied since we can assign each quantum gate in $V$ with a corresponding quantum channel and the composite of these channels forms $\mathcal{C}_V[\cdot]$. 

To satisfy the first requirement, we need $\gamma$ to be relatively large. The largest achievable $\gamma$ is related to the singular values of $S$ which is further related to the Hadamard transform of $|\psi_S\rangle$:
\begin{theorem}[Upper bound of $\gamma$]\label{ubg}
$\gamma$ in $\rho_S$ has the upper bound:
$$\gamma\leq  \gamma_S=\frac{1}{2\sum_\alpha|\langle \alpha|H^{\otimes n}|\psi_S\rangle|} .$$
\end{theorem}
\noindent See Appendix \ref{C} for details. For example, when $|\psi_S\rangle=|0\rangle^{\otimes n}$, we have the smallest $\gamma_S=2^{-n/2-1}$, which corresponds to $\rho_S=2^{-n-1}(I+X)\otimes I_n$; when $|\psi_S\rangle=|+\rangle^{\otimes n}$, we have the largest $\gamma_S=1/2$, which corresponds to $\rho_S=|+\rangle\langle +|^{\otimes (n+1)}$. Since $\gamma=\eta\gamma_{0}$, we can set the initial state $|\psi_{S_0}\rangle$ to have the largest $\gamma_{S_0}$ such as $|+\rangle^{\otimes n}$. 

Since the circuit $V$ is a bridge between $|\psi_S\rangle$ and $|\psi_{S_0}\rangle$ with $\gamma=\eta\gamma_{0}$ and both $\gamma_{0}$ and $\gamma$ have their upper bound, Theorem \ref{ubg} therefore also sets the upper bound of $\eta$. If we have: $\sum_\alpha|\langle \alpha|H^{\otimes n}V|\psi_{S_0}\rangle|=\sum_\alpha|\langle \alpha|H^{\otimes n}|\psi_{S_0}\rangle|$
that holds for any $|\psi_{S_0}\rangle$, $\eta=1$ is allowed, which corresponds to the case where $H^{\otimes n}VH^{\otimes n}$ is composed of permutation circuits and diagonal phase circuits. Since such circuits can be built by $\{S_g,T,\text{CNOT}\}$, the gate $HS_gH$, $HTH$, and $(H\otimes H) \text{CNOT}(H\otimes H)$ in $V$ can be constructed under Pauli computing computing by a quantum channel of the form Eq. \ref{channel} with $\eta=1$. For the Hadamard gate, since $H|+\rangle=|0\rangle$ where $|+\rangle$ has the $\gamma$-upper bound $1/2$ and $|0\rangle$ has the $\gamma$-upper bound $1/(2\sqrt{2})$, the best we can expect for $\eta$ of $HHH=H$ is therefore $1/\sqrt{2}$. We give concrete optimal CBE constructions for $\{H,HS_gH, HTH, (H\otimes H) \text{CNOT}(H\otimes H)\}$ in the Appendix \ref{A1}. 

When $V$ is built by the above 4 gates, it has the form $V=H^{\otimes n}UH^{\otimes n}$ with $U$ a unitary circuit composed of $\{H,S_g,T,\text{CNOT}\}$. Then the amplitude such as $\langle 0 |^{\otimes n}V|\psi_{S_0}\rangle$ with $|\psi_{S_0}\rangle=|+\rangle^{\otimes n}$ can be re-expressed as:
\begin{eqnarray}
\langle 0 |^{\otimes n}V|\psi_{S_0}\rangle=\langle +|^{\otimes n} U|0\rangle^{\otimes n}.
\end{eqnarray}
Combined with previous discussions, the estimation of such amplitudes under Pauli quantum computing has the following theorem:
\begin{theorem}[Estimating $\langle +|^{\otimes n} U|0\rangle^{\otimes n}$]\label{ampesti}
When $U$ is composed of $\{H,S_g,T,\text{CNOT}\}$ with the number of Hadamard gates is $k$ and $V=H^{\otimes n}UH^{\otimes n}$, the (real part and imaginary part of the) amplitude $\langle +|^{\otimes n} U|0\rangle^{\otimes n}$ can be estimated to an additive error $\epsilon$ by Pauli quantum computing with a query complexity:
$$\mathcal{O}(2^{-(n-k)/2}\epsilon^{-1}),$$
on $\mathcal{C}_V$ whose gate complexity is comparable with $U$.
\end{theorem}
In SQC, to estimate such an amplitude, the query complexity on $U$ is $\mathcal{O}(\epsilon^{-1})$. Thus, whenever $k$ is $\mathit{o}(n)$ such as $\text{poly}\log(n)$ and $\sqrt{n}$, we can have an exponential reduction ($2^{-(n-\mathit{o}(n))/2}$) in the complexity of estimation using PQC. This $k=\mathit{o}(n)$ requirement is of particular interest in terms of classical hardness: While it excludes advantages for arbitrary circuits, $\mathit{o}(n)$ Hadamard gates can already create a sub-exponential number of superpositions, which combined with the fact that there are no additional restrictions on the number of $\text{CNOT}$ (entanglement) and $T$ (magic) gates, makes classical simulation algorithms based on tensor network (MPS) \cite{cirac2021matrix} and stabilizers \cite{bravyi2019simulation} inefficient. Therefore, PQC improves quantum computing in a quantum advantage regime.

\subsection{Searching by Pauli searching oracle}
Grover's searching algorithm \cite{grover1996fast} is one of the most important quantum algorithms in quantum computing. In the algorithm, we are given access to the unitary searching oracle:
\begin{eqnarray}
U_{search}=-I_n+2|x\rangle\langle x|,
\end{eqnarray}
which has the property: $U_{search}|x\rangle=|x\rangle$ and $U_{search}|\alpha\rangle=-|\alpha\rangle$ for $\alpha\neq x$. For a $n$-bit searching problem with a unique target, both the query and time complexity of Grover's algorithm on $U_{search}$ to solve the problem is $\mathcal{O}(2^{n/2})$, which gives a quadratic speedup over classical algorithms. 

Imitating the effect of $U_{search}$, in PQC, we consider that we have access to a Pauli searching oracle manifested as a channel:
\begin{eqnarray}\label{pso}
\mathcal{C}_{search}[X\otimes Q_{x}]&&=\eta X\otimes Q_{x},\nonumber\\\mathcal{C}_{search}[X\otimes Q_{\alpha}]&&=-\eta X\otimes Q_{\alpha}\text{, for $\alpha\neq x$}.
\end{eqnarray}
It turns out that quantum channels that can satisfy Eq. \ref{pso} with $\eta=1/3$ is physically realizable as shown in Appendix \ref{B}. It is easy to notice that such an oracle can be used to verify whether an element $\alpha$ is the target by preparing the state $2^{-(n+1)}(I\otimes I_{n}+X\otimes Q_\alpha)$ and feeding it into $\mathcal{C}_{search}[\cdot]$. The expectation value of $X\otimes Q_\alpha$ with respect to the output state is $1/3$ if $\alpha=x$ and $-1/3$ if $\alpha\neq x$. The verification procedure can therefore be done by Pauli measurements in a constant time (assuming the time to implement the oracle is 1) with $\mathcal{O}(1)$ queries on $\mathcal{C}_{search}[\cdot]$.

To use such a channel to solve the search problem, we can further add an ancilla qubit that controls the channel: only when the ancilla qubit is $|1\rangle$, $\mathcal{C}_{search}[\cdot]$ is acted on the system. We use $c-\mathcal{C}_{search}[\cdot]$ to denote this controlled version. Now, starting from an initial state $\rho_{in}$:
\begin{equation}\label{rhoin}
\rho_{in}=\left(\frac{\eta}{1+\eta}|0\rangle\langle 0|+\frac{1}{1+\eta}|1\rangle\langle 1|\right)\otimes |+\rangle\langle+|^{\otimes (n+1)},
\end{equation}
the action of this $\eta=1/3$ $c-\mathcal{C}_{search}[\cdot]$ channel gives (Appendix \ref{B}):
\begin{eqnarray}\label{oott}
&&\rho_{out}=\Tr_{anc}(c-\mathcal{C}_{search}[\rho_{in}])=\nonumber\\&&\frac{1}{2} \begin{pmatrix}\frac{1}{2}|+\rangle\langle +|^{\otimes n}+2^{-(n+1)}I_n& 2^{-(n+1)}Q_x\\2^{-(n+1)} Q_x &\frac{1}{2}|+\rangle\langle +|^{\otimes n}+2^{-(n+1)}I_n\end{pmatrix}.\nonumber\\
\end{eqnarray}
From the explicit form of $\rho_{out}$, we can find that on non-diagonal blocks of $\rho_{out}$, only $Q_x$ encoding the target of the search problem is left with others canceled out. Now regarding the target extraction, if we only care about the query complexity on $\mathcal{C}_{search}[\cdot]$, the strategy is simple. Since the $\{|+\rangle,|-\rangle\}^{\otimes {n+1}}$ basis is the eigenbasis for all $X\otimes Q_\alpha$ and we have $\Tr(X\otimes Q_x)=1/2$, $\Tr(X\otimes Q_\alpha)=-1/2$ for $\alpha\neq x$, we can simply do the $X$-basis measurements for $\mathcal{O}(1)$ times ($\mathcal{O}(1)$ queries on $\mathcal{C}_{search}[\cdot]$) with exponential-time classical post-processing to use the samples to test each $Q_\alpha$ until we find the target.

Interestingly, the special structure of $\rho_{out}$ also enables a measurement strategy to obtain the target information in efficient time complexity if we are given the assurance that $\rho_{out}$ must have the form of Eq. \ref{oott}. We can re-write $\rho_{out}=(\rho_I+\rho_x)/2$ with $\rho_I= I/2\otimes |+\rangle\langle +|^{\otimes n}$ and $\rho_x=2^{-(n+1)}I\otimes I_n+2^{-(n+1)}X\otimes Q_x$. Measuring $\rho_{out}$ on the $X$-basis, the first component $1/2\rho_I$ of $\rho_{out}$ will contribute $1/4$ probability on $ |+\rangle\langle +|\otimes |+\rangle\langle +|^{\otimes n }$ and $1/4$ probability on $ |-\rangle\langle -|\otimes |+\rangle\langle +|^{\otimes n }$. The remaining $1/2$ probability will be contributed by $1/2\rho_x$ and the possible measurement outcomes $|v_{\beta}\rangle=H^{\otimes (n+1)}|\beta\rangle$ with $\beta$ an $n+1$ bit string $\beta=\beta_0\beta_1...\beta_n$ are all $+1$ eigenstates of $X\otimes Q_x$ satisfying:
\begin{equation}\label{ste}
(-1)^{\beta_0+x_1 \beta_1+x_2\beta_2+...x_n \beta_n}=1.
\end{equation}

Due to the symmetry in Eq. \ref{ste}, we can also understand this equation where the measurement outcome $|v_{1x}\rangle$ is the $+1$ eigenstate of $Q_{\beta}$ i.e. $Q_{\beta}$ is a stabilizer of $|v_{1x}\rangle$. Since there are $2^n$ possible outcomes and all of them are stabilizers, they form a stabilizer group with $n$ non-trivial generators. Therefore, we can understand this as an error correction code \cite{gottesman1997stabilizer} with $|+\rangle^{\otimes (n+1)}$ the logical $|\bar{0}\rangle$ and $|v_{1x}\rangle$ the logical $|\bar{1}\rangle$. Thus, to acquire $x$, we simply need to find the logical $\bar{X}$ gate with the property: $H^{\otimes (n+1)}\bar{X}|+\rangle^{\otimes (n+1)} =|1\rangle\otimes |x\rangle$.

To find the logical $\bar{X}$ gate, we need first to do $X$-basis measurements on $\rho_{out}$ to obtain $n$ independent stabilizer generators. To exclude the contribution from $\rho_I$, we abandon the outcomes $ |+\rangle\otimes |+\rangle^{\otimes n}$ and $ |-\rangle\otimes |+\rangle^{\otimes n }$ and only collect the others, which results in an efficiency $1/2$ (with exponentially small disturbance by $\rho_x$). Another efficiency loss is that we cannot ensure that the measurement outcomes are independent generators. The probability that $n$ measurement outcomes are independent is:
\begin{equation}
p_n=\left(1-\frac{1}{2^n}\right)\left(1-\frac{2}{2^n}\right)...\left(1-\frac{2^{n-1}}{2^n}\right)> \frac{1}{4}.
\end{equation}
Thus, the overall efficiency is at least $1/8$, resulting in the overall $\mathcal{O}(n)$ query complexity. After finding $n$ stabilizers $\{Q_{\beta[1]},...,Q_{\beta[n]}\}$, what we need to do is to solve a linear system of equations:
\begin{equation}
\beta[k]_{0}+\sum_{l=1}^n \beta[k]_{l} r_l=0\text{, for $k=1,2,...,n$}.
\end{equation}
This linear system of equations \cite{gottesman1997stabilizer} comes from the fact that the logical $\bar{X}$ gate is commuting with all generators and must be composed of Pauli operators $I$ and $Z$. The solution $r=r_1...r_n$ to the equations will give the logical gate $\bar{X}=Z\otimes Z^{r_1}\otimes...\otimes Z^{r_n} $, acting which on $|+\rangle^{\otimes (n+1)}$ will tell us $x$. Solving this linear system of equations requires a time of order $\mathcal{O}(n^{2.376})$ by the state-of-the-art classical algorithm \cite{coppersmith1987matrix}. This whole efficient target extraction procedure is possible because $I$ and $X$ are commute, which is also the reason why we only choose $I$ and $X$ for encoding and ignore $Y$ and $Z$. We summarize the results in the following theorem:
\begin{theorem}[Searching by Pauli searching oracle]
For the searching problem with a unique target, given access to a Pauli searching oracle $\mathcal{C}_{search}[\cdot]$ manifested as a quantum channel satisfying Eq. \ref{pso} and $\eta=1/3$, and given the assurance that $\rho_{out}$ has the form Eq. \ref{oott}, the target can be found with $\mathcal{O}(n)$ query complexity on $\mathcal{C}_{search}[\cdot]$ and $\mathcal{O}(\text{poly}(n))$ time complexity. 
\end{theorem}
An important open problem here is how efficient it can be to construct such a Pauli searching oracle. In Grover's algorithm, the oracle $U_{search}$ can be efficiently constructed from the description of the problem. For example, given a 3-SAT instance \cite{schaefer1978complexity}, we can translate its conjunctive normal form first into a classical reversible circuit \cite{bennett1988notes} and then a quantum circuit which after a standard trick \cite{preskill1998lecture}, becomes a phase oracle $U_{search}$. However, a similar procedure under PQC may lead to exponentially small $\eta$. While we have given a concrete $\eta=1/3$ construction in Appendix \ref{B}, it only aims to show such an oracle exists but we don't know how to construct it merely from the problem description. Therefore, it is important to explore whether there are novel constructions for Pauli searching oracles with both efficiency and practicability. It is well-known by the BBBV theorem \cite{bennett1997strengths} that the quadratic speedup in Grover's algorithm is already optimal, we don't know whether this also sets the limitation for searching under PQC.

\section{Summary and Outlook}
In summary, we have proposed Pauli quantum computing (PQC) where we treat $I$ and $X$ in non-diagonal blocks of density matrices as $|0\rangle$ and $|1\rangle$ in standard quantum computing. The operations on "quantum states" in PQC including unitary circuits in standard quantum computing can be realized by specially designed quantum channels. We can apply different measurement strategies in PQC to acquire values of operator expectations and amplitudes respectively. We give three examples to help the readers to understand various aspects of PQC. In the first example, we show how to use Lindbladians to implement imaginary time evolution in PQC. We propose the density matrix stabilizer coherence characterizing the steady subspace of a class of Lindbladians, which are stabilizer ground states in Pauli quantum computing. In the second example, we show the advantage of using PQC formalism to improve amplitude estimation. Specifically, we show that for estimating amplitudes $\langle +|^{\otimes n}U|0\rangle^{\otimes n}$ with $U$ containing $\mathit{o}(n)$ Hadamard gates, PQC can exponentially reduce the gate (time) complexity compared with standard quantum computing. In the third example, we show that when we are given access to a Pauli searching oracle in PQC picture mimicking the search oracle in Grover's algorithm, the searching problem can be solved in $\mathcal{O}(n)$ queries and $\mathcal{O}(\text{poly}(n))$ time.

There are several interesting directions that can be further investigated in the future. $\textbf{A}$: In the second example, can we encode solutions of some classical hard problems with practical interest into amplitudes that PQC can give additional advantages over standard quantum computing? $\textbf{B}$: Can we find efficient ways to construct Pauli searching oracles in the third example to give practical quantum speedup and what is the limitation of the approach? $\textbf{C}$: Since the operations in PQC are not necessarily unitary, are there naturally non-unitary problems that PQC can have advantages over standard quantum computing? $\textbf{D}$: Besides treating $I$ and $X$ as $|0\rangle$ and $|1\rangle$, are there other choices of information encoding that can lead to interesting properties? PQC is still in its early stage, we hope this work can invoke explorations along this avenue and give new possibilities for quantum computing and quantum algorithms.

\begin{acknowledgments}
Z.S. would like to thank Wenju Yu, Zihan Chen, Xingjian Zhang, Tianfeng Feng, Jue Xu, and Qi Zhao for fruitful discussions and answering relevant questions. 
\end{acknowledgments}
\bibliography{ref}
\clearpage
\onecolumngrid
\begin{appendix}

\section{Optimal channel constructions of gates in Pauli quantum computing\label{A1}}

Here, we show how to use CBE to construct $\{X,Y,Z,H,HS_gH, HTH, (H\otimes H) \text{CNOT}(H\otimes H)\}$. Basically, when $V$ is one of these gates, we want to find a channel of the form Eq. \ref{channel} such that the requirement Eq. \ref{po} is satisfied. While we have various choices on $F_0$, it turns out that setting it as the projector on the all zero basis state can give optimal $\eta$ for all these gates. For single-qubit gates, the requirement of Eq. \ref{po} is reduced to:
\begin{equation}
\sum_i K_i\otimes L_i^*=\eta\text{CNOT}(|+\rangle\langle+|\otimes V) \text{CNOT}.
\end{equation}

When $V$ is $H$, the 4 Kraus operators of its corresponding channel are:
\begin{equation}
\left\{\begin{pmatrix}
K_i & 0\\
0& L_i
\end{pmatrix}\right\}=\left\{\frac{1}{2}\begin{pmatrix}
I & 0\\
0& X
\end{pmatrix},\frac{1}{2}\begin{pmatrix}
Z & 0\\
0& Z
\end{pmatrix},\frac{1}{2}\begin{pmatrix}
X & 0\\
0& I
\end{pmatrix},\frac{1}{2}\begin{pmatrix}
Y & 0\\
0& Y
\end{pmatrix} \right\}.
\end{equation}
This channel gives the $H$ construction in PQC with optimal $\eta=1/\sqrt{2}$.

When $V$ is $HS_gH$, the 2 Kraus operators of its corresponding channel are:
\begin{equation}
\left\{\begin{pmatrix}
K_i & 0\\
0& L_i
\end{pmatrix}\right\}=\left\{\frac{1}{\sqrt{2}}\begin{pmatrix}
I & 0\\
0& \frac{1}{2}\begin{pmatrix}
1+i & 1-i\\
1-i& 1+i
\end{pmatrix}
\end{pmatrix},\frac{1}{\sqrt{2}}\begin{pmatrix}
X & 0\\
0& \frac{1}{2}\begin{pmatrix}
1-i & 1+i\\
1+i& 1-i
\end{pmatrix}
\end{pmatrix} \right\}.
\end{equation}
This channel gives the $HS_gH$ construction in PQC with optimal $\eta=1$.

When $V$ is $HTH$, the 2 Kraus operators of its corresponding channel are:
\begin{equation}
\left\{\begin{pmatrix}
K_i & 0\\
0& L_i
\end{pmatrix}\right\}=\left\{\frac{1}{\sqrt{2}}\begin{pmatrix}
I & 0\\
0& \frac{1}{2}\begin{pmatrix}
1+e^{i\pi/4} & 1-e^{i\pi/4}\\
1-e^{i\pi/4}& 1+e^{i\pi/4}
\end{pmatrix}
\end{pmatrix},\frac{1}{\sqrt{2}}\begin{pmatrix}
X & 0\\
0& \frac{1}{2}\begin{pmatrix}
1-e^{i\pi/4} & 1+e^{i\pi/4}\\
1+e^{i\pi/4}& 1-e^{i\pi/4}
\end{pmatrix}
\end{pmatrix} \right\}.
\end{equation}
This channel gives the $HTH$ construction in PQC with optimal $\eta=1$.

For the two-qubit gate $(H\otimes H) \text{CNOT}(H\otimes H)$, the requirement of Eq. \ref{po} is reduced to:
\begin{equation}\label{cnot}
\sum_i K_i\otimes L_i^*=\eta(\text{CNOT}_{1,3}\otimes \text{CNOT}_{2,4})
(|++\rangle\langle++|\otimes  (H\otimes H) \text{CNOT}(H\otimes H)) (\text{CNOT}_{1,3}\otimes \text{CNOT}_{2,4})
\end{equation}
By utilizing the fact $\text{CNOT}=|0\rangle\langle 0|\otimes I+|1\rangle\langle1|\otimes X$, we can find a quantum channel with 8 Kraus operators that construct the gate:
\begin{eqnarray}
&&\left\{\begin{pmatrix}
K_i & 0\\
0& L_i
\end{pmatrix}\right\}=\left\{\frac{1}{2\sqrt{2}}\begin{pmatrix}
I\otimes I & 0\\
0& I\otimes I
\end{pmatrix},\frac{1}{2\sqrt{2}}\begin{pmatrix}
I\otimes X & 0\\
0& I\otimes X
\end{pmatrix},\frac{1}{2\sqrt{2}}\begin{pmatrix}
Z\otimes I & 0\\
0& X\otimes I
\end{pmatrix},\frac{1}{2\sqrt{2}}\begin{pmatrix}
Z\otimes X & 0\\
0& X\otimes X
\end{pmatrix},\right.\nonumber\\&&\left. \frac{1}{2\sqrt{2}}\begin{pmatrix}
X\otimes I & 0\\
0& Z\otimes I
\end{pmatrix},\frac{1}{2\sqrt{2}}\begin{pmatrix}
X\otimes X & 0\\
0& Z\otimes X
\end{pmatrix},\frac{1}{2\sqrt{2}}\begin{pmatrix}
Y\otimes I & 0\\
0& -Y\otimes I
\end{pmatrix},\frac{1}{2\sqrt{2}}\begin{pmatrix}
Y\otimes X & 0\\
0& -Y\otimes X
\end{pmatrix}\right\},
\end{eqnarray}
with optimal $\eta=1$.

For Pauli gates, we give two different optimal constructions, the first one is to still choose $F_0=|0\rangle\langle 0|$, the same choice as the other gates. In this one, when $V=X$, the 2 Kraus operators of its corresponding channel are:
\begin{equation}
\left\{\begin{pmatrix}
K_i & 0\\
0& L_i
\end{pmatrix}\right\}=\left\{\frac{1}{\sqrt{2}}\begin{pmatrix}
I & 0\\
0& X
\end{pmatrix},\frac{1}{\sqrt{2}}\begin{pmatrix}
X & 0\\
0& I
\end{pmatrix} \right\}.
\end{equation}
This channel gives the $X$ construction in PQ with optimal $\eta=1$. When $V=Y$, the 2 Kraus operators of its corresponding channel are:
\begin{equation}
\left\{\begin{pmatrix}
K_i & 0\\
0& L_i
\end{pmatrix}\right\}=\left\{\frac{1}{\sqrt{2}}\begin{pmatrix}
I & 0\\
0& Y
\end{pmatrix},\frac{1}{\sqrt{2}}\begin{pmatrix}
Y & 0\\
0& Z
\end{pmatrix} \right\}.
\end{equation}
This channel gives the $Y$ construction in PQC with optimal $\eta=1$. When $V=Z$, the 2 Kraus operators of its corresponding channel are:
\begin{equation}
\left\{\begin{pmatrix}
K_i & 0\\
0& L_i
\end{pmatrix}\right\}=\left\{\frac{1}{\sqrt{2}}\begin{pmatrix}
I & 0\\
0& Z
\end{pmatrix},\frac{1}{\sqrt{2}}\begin{pmatrix}
Y & 0\\
0& Y
\end{pmatrix} \right\}.
\end{equation}
This channel gives the $Z$ construction in PQC with optimal $\eta=1$. The second optimal ($\eta=1$) construction that is more natural and used in Section \ref{example1} is to choose $F_0=I$, which leads to:
\begin{eqnarray}
&&\text{$X$: }\begin{pmatrix}
K & 0\\
0& L
\end{pmatrix}=\begin{pmatrix}
I & 0\\
0& X
\end{pmatrix},\nonumber\\&&\text{$Y$: }\begin{pmatrix}
K & 0\\
0& L
\end{pmatrix}=\begin{pmatrix}
Z & 0\\
0& -Y
\end{pmatrix},\nonumber\\&&\text{$Z$: }\begin{pmatrix}
K & 0\\
0& L
\end{pmatrix}=\begin{pmatrix}
Z & 0\\
0& Z
\end{pmatrix}.
\end{eqnarray}

\section{Proof of Theorem \ref{ubg}\label{C}}
Under the mapping $\mathcal{PQC}[\cdot]$, we have:
\begin{eqnarray}
\mathcal{PQC}\left[I/\sqrt{2}\right]&&=|0\rangle,\nonumber\\
\mathcal{PQC}\left[X/\sqrt{2}\right]&&=|1\rangle,\nonumber\\
\mathcal{PQC}[|+\rangle\langle +|]&&=\mathcal{PQC}[(I+X)/2]=|+\rangle,\nonumber\\
\mathcal{PQC}[|-\rangle\langle -|]&&=\mathcal{PQC}[(I-X)/2]=|+\rangle.
\end{eqnarray}
Therefore, combined with $2^{-n/2}\Tr(Q_\alpha S)=\langle \alpha|\psi_S\rangle$, we have the relation:
\begin{eqnarray}\label{bbb2}
\sum_\alpha|\langle \alpha|H^{\otimes n}|\psi_S\rangle|=\sum_\alpha |\Tr(H^{\otimes n}|\alpha\rangle\langle\alpha|H^{\otimes n}S)| =\Tr(|H^{\otimes n}S H^{\otimes n}|).
\end{eqnarray}
Since $S$ is composed of Pauli $I$ and $X$ operators, $\Sigma_S=H^{\otimes n}S H^{\otimes n}$ is diagonal. Under this diagonal basis, we have:
\begin{eqnarray}
(I\otimes H^{\otimes n} )\rho_S (I\otimes H^{\otimes n} )=\begin{pmatrix}
D_0 & \gamma \Sigma_S \\
\gamma \Sigma_S^* &D_1
\end{pmatrix}.
\end{eqnarray}
with $D_0$ and $D_1$ two unnormalized density matrices satisfying $\Tr(D_0+D_1)=1$. Now, if $\Sigma_S=\sum_\alpha \chi_\alpha |\alpha\rangle\langle \alpha|$, then we must have:
\begin{eqnarray}\label{ppppp}
\gamma \sum_\alpha |\chi_\alpha|\leq \gamma\sum_\alpha \sqrt{p_{0,\alpha}p_{1,\alpha}}\leq \sum_\alpha \frac{p_{0,\alpha}+p_{1,\alpha}}{2}=\frac{1}{2},
\end{eqnarray}
where $p_{0,\alpha}$ and $p_{1,\alpha}$ are diagonal values of $D_0$ and $D_1$ respectively. The first inequality in Eq. \ref{ppppp} is because $(I\otimes H^{\otimes n} )\rho_S (I\otimes H^{\otimes n} )$ is a density matrix. Combining Eq. \ref{ppppp} with Eq. \ref{bbb2}, Theorem \ref{ubg} is proved.

\section{Existence of $\eta=1/3$ Pauli searching oracle\label{B}}
Essentially, we want to show that the Pauli searching oracle in Eq. \ref{pso} can be physically realized by a quantum channel of the form Eq. \ref{channel} satisfying Eq. \ref{po} with $V=U_{search}$. By setting $F_0=(|0\rangle\langle 0|)^{\otimes n}$, Eq. \ref{po} is reduced to:
\begin{eqnarray}
\sum_i K_i\otimes L_i^*=\eta (\text{CNOT}_{1,n+1}\otimes...\otimes\text{CNOT}_{n,2n}) ((|+\rangle\langle +|)^{\otimes n}\otimes U_{search})(\text{CNOT}_{1,n+1}\otimes...\otimes\text{CNOT}_{n,2n}) .
\end{eqnarray}
In the following, we will use $\text{CNOT}_{2n}$ to denote $(\text{CNOT}_{1,n+1}\otimes...\otimes\text{CNOT}_{n,2n})$. Since $U_{search}=-I_n+2|x\rangle\langle x|$, we can consider $-I_n$ and $2|x\rangle\langle x|$ separately. For $-I_n$, we have:
\begin{eqnarray}\label{in}
&&\text{CNOT}_{2n}((|+\rangle\langle +|)^{\otimes n}\otimes -I_n)\text{CNOT}_{2n}\nonumber\\=&& -\frac{1}{2^n}(I\otimes I+X\otimes X)_{1,n+1}\otimes...\otimes(I\otimes I+X\otimes X)_{n,2n}\nonumber\\=&& \frac{1}{2^n}\sum_{i=0}^{2^n-1} Q_i\otimes -Q_i,
\end{eqnarray}
where $(I\otimes I+X\otimes X)_{1,n+1}$ indicates Pauli operators on the 1st and $n+1$th qubits, which can be realized by a quantum channel $\mathcal{C}_I[\cdot]$ with $2^n$ Kraus operators:
\begin{equation}
\begin{pmatrix}
K_i & 0\\
0& L_i
\end{pmatrix}=\frac{1}{2^{n/2}}\begin{pmatrix}
Q_i & 0\\
0& -Q_i
\end{pmatrix}\text{, for each i in $\{0,1\}^{\otimes n}$}.
\end{equation}
For $2|x\rangle\langle x|$, we have:
\begin{eqnarray}\label{xx}
&&\text{CNOT}_{2n}((|+\rangle\langle +|)^{\otimes n}\otimes |x\rangle\langle x|)\text{CNOT}_{2n}\nonumber\\=&& \frac{1}{2^n}(|0\rangle\langle 0|\otimes|0\oplus x_1\rangle\langle 0\oplus x_1| +|0\rangle\langle 1|\otimes|0\oplus x_1\rangle\langle 1\oplus x_1|\nonumber\\&&+|1\rangle\langle 0|\otimes|1\oplus x_1\rangle\langle 0\oplus x_1|+|1\rangle\langle 1|\otimes|1\oplus x_1\rangle\langle 1\oplus x_1|)_{1,n+1}\otimes...\otimes(\cdot)_{n,2n}\nonumber\\=&& \frac{1}{2^n}\sum_{i,j=0}^{2^n-1} |i\rangle\langle j|\otimes |i\oplus x\rangle\langle j\oplus x|
\end{eqnarray}
where $x=x_1...x_n$ is its binary representation, which can be realized by a quantum channel $\mathcal{C}_x[\cdot]$ with $2^{2n}$ Kraus operators:
\begin{equation}
\begin{pmatrix}
K_{i,j} & 0\\
0& L_{i,j}
\end{pmatrix}=\frac{1}{2^{n/2}}\begin{pmatrix}
|i\rangle\langle j| & 0\\
0& |i\oplus x\rangle\langle j\oplus x|
\end{pmatrix}\text{, for each $i,j$ in $\{0,1\}^{\otimes n}$}.
\end{equation}
Thus, to realize $V=U_{search}$, we can simply let:
\begin{equation}
\mathcal{C}_{search}[\cdot]=\frac{2}{3}\mathcal{C}_x[\cdot]+\frac{1}{3}\mathcal{C}_I[\cdot],
\end{equation}
which gives $\eta=1/3$. Acting the controlled version $c-\mathcal{C}_{search}[\cdot]$ to $\rho_{in}$ of Eq. \ref{rhoin} gives:
\begin{eqnarray}
\rho_{out}=\Tr_{anc}(c-\mathcal{C}_{search}[\rho_{in}])&&=\frac{1}{8}\begin{pmatrix}|+\rangle\langle +|^{\otimes n}& |+\rangle\langle +|^{\otimes n}\\|+\rangle\langle +|^{\otimes n }&|+\rangle\langle +|^{\otimes n}\end{pmatrix}+\frac{3}{8}\begin{pmatrix}\frac{1}{3}|+\rangle\langle +|^{\otimes n}+\frac{2}{3}2^{-n}I_n & \frac{1}{3}2^{-n}\left(Q_x-\sum_{\alpha\neq x}Q_\alpha\right)\\\frac{1}{3}2^{-n}\left(Q_x-\sum_{\alpha\neq x}Q_\alpha\right)&\frac{1}{3}|+\rangle\langle +|^{\otimes n}+\frac{2}{3}2^{-n}I_n \end{pmatrix}\nonumber\\&&=
\frac{1}{2} \begin{pmatrix}\frac{1}{2}|+\rangle\langle +|^{\otimes n}+\frac{1}{2}2^{-n}I_n& \frac{1}{2} 2^{-n}Q_x\\\frac{1}{2} 2^{-n}Q_x &\frac{1}{2}|+\rangle\langle +|^{\otimes n}+\frac{1}{2}2^{-n}I_n\end{pmatrix}.
\end{eqnarray}
\end{appendix}
\end{document}